\documentstyle[aps]{revtex}
\begin{document}
\draft

\title{Is Compton scattering in magnetic fields really infrared divergent?%
        \footnote{To appear in Phys. Rev. {\bf D51} (2), 1995.}}

\author{M. Kachelrie{\ss}, D. Berg, and G. Wunner}
\address{Theoretische Physik I, Ruhr-Universit\"at Bochum,
         D-44780 Bochum, Germany}

\date{Received 31 August 1994}
\maketitle

\begin{abstract}
 The infrared behavior of QED changes drastically in the presence of
 a strong magnetic field: the electron self-energy and the vertex function
 are infrared {\em finite}, in contrast with field-free QED, while
 new infrared divergences   appear that are absent in free
 space. One famous example of the latter is the infrared catastrophe of
magnetic
 Compton scattering, where the cross section for scattering of
 photons from electrons which  undergo a
 transition  to the Landau ground state {\em diverges} as the
 frequency of the incoming photon goes to zero.
 We examine this divergence in more detail
 and prove that  the singularity of the cross section is
 {\em removed} as soon as proper
 account is taken of all   quantum electrodynamical processes that become
 indistinguishable from  Compton  scattering in the limit of vanishing
 frequency of the incident  photon.
\end{abstract}


\section{INTRODUCTION}

Compton scattering is the central mechanism for the redistribution of energy
of hot electrons and electromagnetic radiation, and thus for the  formation
of spectra, in strongly magnetized pulsating x-ray sources \cite{Tr86}
and, possibly, certain magnetized $\gamma$-ray burst sources \cite{Ha91}.
Therefore, this quantum electrodynamical
process was among the first to be recalculated for the magnetic field strengths
of several $10^{8}$ T  which were  detected in these objects
by way of identification of cyclotron line features. The recalculation was
necessitated by the fact that at these field strengths the cyclotron energy
becomes of the
order  of the electron rest energy (equality holds
for  $B =   m_{\rm e}^2 /e  = 4.414 \times 10^{9}$ T),
 and thus, in  calculating quantum electrodynamical
processes,  the quantization of the electron states into discrete Landau
levels has to  be  fully taken into account \cite{Ca77}. This leads to
a drastic change in the   structure of cross sections as compared to free
space.
The relativistic Compton scattering
cross section in a strong magnetic field was  first derived by Herold
\cite{He79}
 for
initial and final electrons in the Landau ground state ($n_i = n_f = 0$, where
$n$ denotes the Landau quantum number), by Melrose and Parle \cite{Me83} for
 $n_i=0$ and
final states $n_f=0, 1$, by Daugherty and Harding \cite{Da86}
for $n_i=0$ and
arbitrary final states, and Bussard {\it et al.} \cite{Bu86} for
arbitrary initial and final states.

Because of the mathematical complexity of the resulting
expressions, the numerical evaluations and implementations
into actual radiative transfer calculations   were first    restricted   to
considering Compton scattering with  electrons in the Landau ground state
\cite{Ha89}. It came therefore
as a surprise when Brainerd  \cite{Br89} pointed out,
for the special case $n_i=1$,
$n_f=0$, that the cross section for  Compton
   scattering with  electrons  in {\it excited}\/  Landau states
which during the scattering process   undergo a transition
to the Landau ground state
{\it diverges}\/
as the  energy  of the incoming photon goes to zero. Obviously, this
process very efficiently turns soft  photons into scattered
cyclotron photons of several tens of keV, and thus
it was  argued that the infrared catastrophe of the Compton cross section
in a magnetic field is at the root of the observed deficiency of soft photons
in the spectra of magnetized $\gamma$-ray burst sources.
It is the purpose of this paper  to  examine the divergence more closely.
Our main result is  that the singularity of the cross section
is removed when proper
account is taken of all the quantum electrodynamical processes which
degenerate with Compton  scattering in the limit of vanishing frequency of
the incoming photon.

Let us first briefly recall what is  the cause of the infrared catastrophe
in ordinary quantum electrodynamics (cf. \cite{Ja76,Ki} and references
therein). It is known since the classic paper of Bloch and Nordsieck
\cite{Bl37} that  infrared divergences appear in   theory because, loosely
speaking,  an accelerated
charged particle can emit an infinite number of soft photons with finite
total energy. In the real world, any experiment is carried out during a finite
time interval, so a finite energy resolution $\Omega$ is necessarily inherent
in every experiment (a lower bound is   given by the energy
uncertainty, $h/\Delta T$,
although the energy resolution $\Omega$ of a real detector will in general
be much larger).
Therefore, whenever charged particles participate in  some reaction, one cannot
distinguish experimentally between this reaction and the same one with
supplementary (real or virtual) soft photons being emitted or absorbed. Here
the term ``soft photons''
means photons which are undetectable because their   energies lie below
the detection threshold,  $\omega_s < \Omega$
(throughout this paper the subscript $s$ stands for soft photons).
For that reason, one way to solve the infrared problem is to take
into account
in the calculation of transition probabilities
the coherent superposition of the $S$-matrix elements of all
indistinguishable processes, in  accordance with   the general principles
of quantum theory \cite{Fe}.
Then, as   is well known from  field-free QED,
the infrared divergent contributions to observables
should cancel in any order of perturbation theory.
However, the mechanism of cancellation has to be quite different
in QED in magnetic fields: e.g., in Ref. \cite{Co72} it was proved that the
electron self-energy in a magnetic field is {\em not} infrared divergent.
On the other hand, there occurs a new infrared divergence in the magnetic
Compton
cross section for incident photons. Although this cross section is finite for
every finite energy $\omega_i$ of the incident photon, its "explosion" for
$\omega_i \to 0$ seems unphysical. We note that
in a similiar case of infrared divergences for incident photons
in the older theory of weak and electromagnetic interactions
their cancellations were shown in Ref. \cite{Nucl}.

Our line of argument can be illustrated most easily with the help of the
Feynman diagrams shown in Fig.~1.
In the limit of vanishing frequency  $\omega_i$ of the
incident photon, the  second-order process of Compton scattering  from
an Landau excited electron
(Fig.~1.2, $S \propto e^2$) becomes indistinguishable from the process of
cyclotron emission  (Fig.~1.1, $S \propto e^1$). Thus,  in
calculating the transition probability in this limit,
the two corresponding
$S$-matrix elements must first be added coherently and subsequently be squared.
The squared total $S$-matrix element then contains terms up to
order  $e^4$, which implies that,  for the expansion to be
consistent,  in the coherent
superposition all other processes  must also be included whose direct or
cross terms  produce
contributions up to order $e^4$ in the squared total $S$-matrix element,
and degenerate with cyclotron emission
in the limit of one or more of the photon frequencies involved going to
zero.  Obviously these are the following second- and third-order processes:
double cyclotron emission  with one soft photon (Fig.~1.3), triple
cyclotron emission  with two  soft photons (Fig.~1.7), double Compton
scattering
with two soft photons (Fig.~1.5),  (Fig.~1.6), and the low-energy part of the
vertex correction of cyclotron emission (Fig.~1.4).
Note that since  the electron self-energy in a strong magnetic field
is infrared finite we    need not
consider soft photon insertions into the external electron lines \cite{Note}.

In what follows, we will not calculate the infrared finite part of the
magnetic Compton cross section but restrict ourselves to the simpler
task of investigating the singular terms of the $S$-matrix elements
and demonstrate that a cancellation of their divergences occurs.
It is therefore sufficient to
consider the $S$-matrix elements in the limit $\omega_s \to 0$.

We shall prove below
that  in the limit $\omega_s \to 0$
the contributions
of (ordinary) Compton scattering and double cyclotron emission
to the total $S$-matrix element {\em cancel},
as do the contributions of   double Compton scattering, while the
term  due to triple cyclotron emission vanishes in the cross section,
and the vertex correction
remains finite. Thus    the infrared catastrophe of
the Compton scattering  cross section in magnetic fields reported in the
literature is {\em nonexistent}. We note, however,  that  $(n_i \ne 0) \to 0$
cross sections may still remain large  compared to the
$0 \to 0$  cross section
at low energies, in which case the essence of Brainerd's \cite{Br89}
analysis would remain valid inspite of  the
absence of a  real singularity of the cross section for vanishing
photon energies.

\section{INFRARED BEHAVIOR OF THE $S$-MATRIX ELEMENTS}

To render  our argument quantitative we start from  the $S$-matrix element
of Compton scattering in a magnetic field, $S_{2}^{(2)}$ (in what follows
the superscript of $S$ denotes the order in $e$, and the subscript the
subcaption number given to the   process in  Fig.~1), which reads
(cf. Bussard {\it et al.} \cite{Bu86})
\begin{eqnarray}   \label{compton}
& S_{2}^{(2)}  =
 \left( \frac{2\pi}{V}\right)^2 \, \frac{e^2}{(\omega_i\omega_f)^{1/2}}
 \;\delta (E_f +\omega_f -E_i -\omega_i ) &
\nonumber\\ &  \times
 \sum_{a,\lambda}
 \left( \frac{\left(\vec\epsilon_f^{\;\ast}\cdot\vec J_{f,a}^{\:(\lambda)}
              \right)
            \left(\vec\epsilon_i \cdot\vec J_{i,a}^{\:(\lambda)\ast}\right)}
             {E_i +\omega_i - \lambda (E_a -i\epsilon_a) }
 + \frac{\left(\vec\epsilon_f^{\;\ast}\cdot\vec J_{i,a}^{\:(\lambda)}\right)
            \left(\vec\epsilon_i \cdot\vec J_{f,a}^{\:(\lambda)\ast}\right)}
             {E_i -\omega_f - \lambda (E_a -i\epsilon_a) } \right) \; .
\end{eqnarray}
Here, $i$ and $f$ refer to the initial and final states, the sum over $a$
runs over the intermediate Landau states
of electrons ($\lambda = + 1$) and positrons ($\lambda = - 1$),
and the quantities  $\vec J$ denote matrix elements
whose explicit forms are given  in \cite{Bu86}. We note that
in the limit $\omega_i \to 0$ these quantities assume  constant, finite values.

The imaginary parts of the energies of the intermediate states, $i\epsilon_a$,
in Eq. (\ref{compton}) account \cite{Lo52} for the fact that  excited
Landau states have nonzero   widths, i.e.,
$i\epsilon_a$ is  given   by
 $\frac{1}{2}i\Gamma_n$, where $\Gamma_n$ is the cyclotron
decay rate  of an electron in the $n$th Landau level.
Obviously, resonances appear in Eq. (\ref{compton}) at the zeros of the real
parts of the energy denominators, which is the case   when
Compton scattering  degenerates into electron cyclotron absorption
($E_i+\omega_i = E_a = E_f$, $\omega_f = 0$) or emission
($E_i-\omega_f = E_a = E_f$, $\omega_i = 0$), i.e.,  the virtual electron is
created ``on-shell''. Because of the nonvanishing widths $\frac{1}{2}i\Gamma_n$
for $n > 0$, these resonances remain finite, with the   exception of
the case of cyclotron transitions to the Landau ground state
($n_a =n_f  =0$, $\omega_i = 0$, $\omega_f \ne 0$): the latter is stable,
viz. $\Gamma_0 =0$, and thus  a genuine singularity occurs in the
expression (\ref{compton}). This is the type of infrared
divergence  pointed out by Brainerd \cite{Br89}.
The infrared divergent part of the $S$-matrix element  is diagramatically
shown in Fig.~2. It can be read off Eq. (\ref{compton}) that the
divergence of $S$   is of the order ${\cal O}(\omega_i^{-3/2})$.

We now turn to the  $S$-matrix element
of double-cyclotron emission, $S_{3}^{(2)}$, which can easily be obtained
from that of Compton scattering using the crossing symmetry replacements
\begin{equation}
 k_{i,s}^{\mu} \to -k^{\mu}_{f,s}  \; .
\end{equation}
In the limit $\omega_{f,s}\to 0$ the terms containing
$\vec k_{f,s}$ reduce to   nondivergent expressions  identical
to those  of $S_{2}^{(2)}$. Because of the replacements
$\omega_{i,s}\to -\omega_{f,s}$ in the energy denominators   it then follows
\begin{equation}
 S_{2}^{(2)} \left(\vec k_s \right) = -S_{3}^{(2)} \left(\vec k_s \right) \; .
\end{equation}
This implies that in the limit $\omega_s \to 0$ the divergences of
Compton scattering and double cyclotron emission identically cancel.

We now have to show  that all other  processes which also  have to be
taken into account produce no new divergences.
This is a simple task for the two third-order processes 5 and 6 in Fig.~1
with at least one soft photon, for which an analogous application of the
crossing symmetry argument  given above yields
\begin{equation}
 S_{5}^{(3)}\left(\vec k_s \right) = -S_{6}^{(3)}\left(\vec k_s \right)  \; ,
\end{equation}
while in the cross section of the process $S_7^{(3)}$ all possible infrared
divergences are canceled by the the phase space factors $d^3 k$.

Thus the only critical term  that remains is the vertex correction to
cyclotron emission (process 4 in Fig.~1). Using the same technique
as described in   \cite{Me94} we have derived,
to our knowledge for the first time,   the vertex  correction in a strong
magnetic field
\cite{Ka93}, but  will restrict ourselves here to a
discussion of the $S$-matrix element only in so far as is necessary to prove
that this process is not infrared divergent.   (A full treatment of the
vertex  correction in a strong magnetic field will be presented elsewhere.)
The $S$-matrix element reads
\begin{eqnarray}
 S^{(3)}_{4} \! & = & \! (ie)^3 \! \int d^4 x \, d^4 x^{\prime} \,
 d^4 x^{\prime\prime} \,
 \bar\psi_f^{(\lambda =+)}(x)\gamma^{\mu}iS_F(x,x^{\prime})
\nonumber\\ \! & \times & \! \gamma^{\nu}
 iS_F(x^{\prime},x^{\prime\prime})\gamma^{\varrho}
 \psi_i^{(\lambda =+)} (x^{\prime\prime})
 iD_{\mu\varrho}(x-x^{\prime\prime})
     A_{\nu}^{\ast}(x^{\prime})  \; ,
\end{eqnarray}
where, as usual, $iD_{\mu\nu}$ denotes the photon propagator, while $iS_F$
describe the electron propagator and $\psi^{(\lambda =\pm)}$
 the electron and positron fields in a magnetic field, respectively.
Using the temporal gauge for the photon propagator and performing
the   integrations over time  we obtain
\begin{eqnarray}
\lefteqn{S^{(3)}_{4}=(-ie)^3 \: 2\pi \, \delta (E_f +\omega_f -E_i) \:
    \int \frac{d^4 k}{(2\pi)^4}}
\nonumber\\ &&
    \int d^3 x \; d^3 x^{\prime} \, d^3 x^{\prime\prime} \,
    \psi_f^{(\lambda =+)\dag}(\vec x) \alpha_j \sum_{a,b}
\nonumber\\ & & \Bigg\{ \bigg(
         \frac{1}{E_i -E_a -\omega -\omega_f +i \epsilon_a} \;
         \frac{1}{E_i -E_b -\omega +i \epsilon_b}
\nonumber\\ &&
  \psi_a^{(\lambda =+)}(\vec x)\psi_a^{(\lambda =+)\dag}(\vec x^{\prime})
  \alpha_k
  \psi_b^{(\lambda =+)}(\vec x^{\prime})
  \psi_b^{(\lambda =+)\dag}(\vec x^{\prime\prime}) \bigg)
\nonumber\\ && + \bigg(
         \frac{1}{E_i -E_a -\omega -\omega_f +i \epsilon_a} \;
         \frac{1}{E_i +E_b -\omega -i \epsilon_b}
\nonumber\\
& & \psi_a^{(\lambda =+)}(\vec x)
    \psi_a^{(\lambda =+)\dag}(\vec x^{\prime})
    \alpha_k
    \psi_b^{(\lambda =-)}(\vec x^{\prime})
    \psi_b^{(\lambda =-)\dag}(\vec x^{\prime\prime}) \bigg)
\nonumber\\ && + \bigg(
         \frac{1}{E_i +E_a -\omega -\omega_f -i \epsilon_a} \;
         \frac{1}{E_i -E_b -\omega +i \epsilon_b}
\nonumber\\
& &  \psi_a^{(\lambda =-)}(\vec x)
     \psi_a^{(\lambda =-)\dag}(\vec x^{\prime})
     \alpha_k
     \psi_b^{(\lambda =+)}(\vec x^{\prime})
     \psi_b^{(\lambda =+)\dag}(\vec x^{\prime\prime}) \bigg)
\nonumber\\ && + \bigg(
         \frac{1}{E_i +E_a -\omega -\omega_f -i \epsilon_a} \;
         \frac{1}{E_i +E_b -\omega -i \epsilon_b}
\nonumber\\
& &  \psi_a^{(\lambda =-)}(\vec x)
     \psi_a^{(\lambda =-)\dag}(\vec x^{\prime})
     \alpha_k
     \psi_b^{(\lambda =-)}(\vec x^{\prime})
     \psi_b^{(\lambda =-)\dag}(\vec x^{\prime\prime}) \bigg) \Bigg\}
\nonumber\\ &&
 \alpha_l \psi_i^{(\lambda =+)}(x^{\prime\prime})
\frac{\epsilon^{\ast}_k}{\sqrt{2\omega_f V}} \;
    e^{-i\vec k_f \vec x^{\prime}}
    iD_{jl}(k)
    e^{+i\vec k (\vec x -\vec x^{\prime\prime})}   \; .
\end{eqnarray}
The spatial integrals lead to  complicated expressions consisting essentially
of polynomials in the momentum of the virtual photon,  they   therefore
contribute only constant terms when the energy of the virtual photon approaches
zero.  The potentially infrared divergent part
of $S_4^{(3)}$ is diagrammatically shown in Fig.~3.
Replacing $i\epsilon_{a,b}$ with $ \frac{1}{2}i\Gamma_n$ as before, restricting
the virtual photon momentum to a domain $\Omega$ defined by the condition
that all processes in point are observationally indistinguishable
and taking into account   $n_i > 0$, we obtain
\begin{equation}
 S^{(3)}_{4} \propto
 \int_{\Omega} d^4 k \;\frac{1}{k^2} \;\frac{1}{\Gamma_{n_i}}\;\frac{1}{\omega}
 \propto \int_{0\leq k \leq\Omega} k^3 dk \;\frac{1}{k^3}<\infty\; ,
\end{equation}
from which it follows that the contribution of low-energy virtual photons
to $S_{4}^{(3)}$ is  {\em not infrared divergent}.

\section{SUMMARY}

In ordinary  quantum electrodynamics   infrared divergences arise in
perturbation theory when  corrections by emitted or virtual soft photons
are taken into account. Examples for soft photon corrections are  the
self-energy and the vertex function.
By contrast,  in QED in strong magnetic fields the self-energy
as well as the vertex function are not infrared divergent, but there
occur  new divergences for {\em incident} soft photons.
For the case of Compton scattering with electrons in excited Landau states
which undergo a transition to the Landau ground state we have shown that
the divergent parts of the observationally indistinguishable processes
cancel for vanishing energy of the soft photons
\begin{equation} \label{zero}
\lim_{\vec k_s \to 0} \;
 \sum_{i\in \{2,3,5,6 \}}
  S_{i} \left(\vec k_s \right)\ = 0 \; ,
\end{equation}
while the contributions of the remaining processes to the
transition probability remain finite or vanish.
Thus there is no infrared catastrophe of the Compton cross section
for $(n_i \ne 0) \to 0$ transitions and, in contrast  to field-free QED,
already
the total $S$-matrix element is infrared finite.

We have derived Eq. (\ref{zero})
using the very general argument of crossing symmetry. This symmetry implies
that a cancellation of the
infrared divergences of emitted and absorbed soft photons will take place in
every order of perturbation theory.

Although we have shown that in the limit $\omega_i \to 0$ the Compton
cross section tends to a finite value, it is difficult,
because of the complexity of the individual $S$-matrix elements,
to answer the question as to the
{\em behavior} of the  cross section as a function of photon energy
in the vicinity of   $\omega_i = 0$.  Very fundamental considerations of
field-free quantum electrodynamics \cite{Ja76} suggest that
for $\omega_i < \Omega \ll m_e$
the basic process without any soft photon -- in our case
first-order cyclotron emission  -- will give the dominant
contribution to the total transition probability $w$, viz.
\begin{equation} \label{cycl}
 w = \frac{1}{T}\; \left| \: S_1^{(1)} \, \right|^2 \; ,
\end{equation}
in which case soft photons  assume the  role of   spectators.
As a consequence the transition probability will  assume the constant
value following
from Eq. (\ref{cycl}) for $\omega_i < \Omega$, which through $\Omega$ depends
on
the actual observational resolution. Therefore, theoretical calculations
of spectra (i.e. Monte-Carlo simulations) which include transitions
$(n_i \neq 0) \to 0$ should be carried out taking  into account the finite
resolution of the specific detector.

\acknowledgments

We are indebted to  Markus Mentzel for helpful discussions  and
Martin Kaiser for his criticism of an earlier view of this problem.
This work was supported in part by Deutsche Forschungsgemeinschaft.
One of us (M.K.) acknowledges a grant by Deutscher Akademischer
Austauschdienst during the final stage of the preparation of this paper.

\newpage

\begin{figure}
\caption{Feynman diagrams of the  processes which become  experimentally
  indistinguishable as the energies of the soft photons $s$   go  to zero.
 (Exchange diagrams have been omitted for brevity.)}
\end{figure}

\begin{figure}
\caption{Infrared divergent part of Compton scattering.}
\end{figure}

\begin{figure}
\caption{Potentially infrared divergent part of the vertex correction
         to cyclotron emission.}
\end{figure}


\begin{references}
\bibitem{Note}
To avoid misunderstandings we emphasize that taking into account all the
processes of Fig.~1 in calculating the magnetic Compton cross section
is essential, and correct, {\sl only} in the limit $\omega_i \to 0$ (and
$\omega_s \to 0$),
and not for arbitrary frequencies $\omega_i$  and $\omega_s$ where the
processes can be discriminated experimentally.
\end{references}
\end{document}